# Effect of charge carrier relaxation during hopping process on electroluminescence in organic solids

Arunandan Kumar[*], Priyanka Tyagi, Ritu Srivastava, M.N. Kamalasanan

*Center for Organic Electronics, National Physical Laboratory (Council of Scientific and Industrial Research), Dr. K.S.Krishnan Road, New Delhi-110012, India*

Energetic disorder in disordered organic solids has been found to alter their physical parameters. Here, we have demonstrated, by means of Monte-Carlo simulation and experiments, that the electroluminescence (EL) spectrum is dependent on energetic disorder. This dependence has been attributed to the charge carrier relaxation during hopping process. The dependence of EL spectrum on energetic disorder makes it temperature dependent and temperature dependence has been found to vary with energetic disorder in a variety of materials. The simulation has been performed by taking the relaxation of charge carriers via transport energy in the Gaussian density of states. An analytical equation was established for spectral shift as a function of transport energy.

---

[*] corresponding author
Email: kumar.arunandan@gmail.com

Since the first report of efficient electroluminescence (EL) in thin film of $Alq_3$ has appeared in 1987[1], there has been significant advancement in the research activity related to organic solids. These efforts resulted in manifold improvement in efficiency of EL devices[2,3]. Despite the technological advancement in organic EL devices, there is a lack of theoretical understanding of the physical process involved in these devices like injection of charge carriers from metal into a film of organic solid[4-7], transport of charge carriers in the organic films[8-17] and the recombination of oppositely charged carriers to form exciton[4], the decay of which giving rise to organic EL.

The studies on EL from anthracene crystal, powder, sublimation flake by William et.al[18] show that the EL spectrum is different for different forms of anthracene. Their study show that the peak of EL spectrum exactly coincide with the optical band gap for anthracene crystal, while for other forms the peak is red shifted in comparison to the EL in the crystal. Further the studies on small molecular organic solids show that the peak of EL spectrum corresponds to an energy which is less than the optical band gap. And this difference varies for different organic solids. The situation become even more unclear when there is a large shift in EL peak for the materials having same optical band gap. These results indicate towards the dependence of EL spectrum on energetic disorder and further studies are required to understand this dependence. The energetic disorder in organic solids affects the physical properties of these materials. In this letter, EL spectrum has been simulated using Monte-Carlo simulation technique to study the effect of energetic disorder on EL from organic solids. The simulation technique includes the relaxation of charge carriers via transport energy during the hopping process. Devices of different organic materials lithium quinolate (Liq), tris (8-hydroxyquinolinato) aluminum ($Alq_3$), [(2-(2-hydroxyphenyl)benzoxazole)(2-methyl-8-hydoxyquinoline)] zinc (Zn(hpb)mq) have been fabricated to further study the effect of energetic disorder on their EL spectrum by measuring the EL spectrum at different temperatures.

The EL spectrum of an OLED has been simulated by recombining a positive charge carrier with a negative charge carrier in the emitting layer. The emitted photon has energy equal to the difference of energies of negative and positive charge carriers. Inside the organic semiconductor, HOMO and LUMO are distributed in a Gaussian DOS [g(E)] with variance $\sigma_{HOMO}$ and $\sigma_{LUMO}$ and centered around $E_{HOMO}$ and $E_{LUMO}$ where g(E) is

$$g(E) = \frac{N}{\sqrt{2\pi}\sigma} \exp\left(-\frac{E^2}{\sigma^2}\right), \qquad - (1)$$

$E_{LUMO}$ and $E_{HOMO}$ are the energies of LUMO and HOMO for which the DOS of energy is maximum.

Movement of charge carriers in organic solids is analogous to the random walk in a rough energy landscape with long range coulomb potential. In this study, we have used Monte-Carlo simulation technique to model charge carrier movement and recombination of electrons and holes inside the organic semiconductor. A test sample of lattice sites 70x70xZ (Z is the thickness of emissive layer divided by lattice constant a) has been generated by computer. Two energy levels have been assigned to each site in which one corresponds to LUMO and other for HOMO. These energies are chosen randomly from a Gaussian shaped DOS distribution of variable width $\sigma_{LUMO}$ and $\sigma_{HOMO}$ corresponding to LUMO and HOMO energies. To avoid the surface effects, periodic boundary conditions were applied in x and y direction. To simulate charge carrier recombination, 10,000 electrons and 10,000 holes were started from z=1a plane, z=Za plane respectively and they were allowed to hop under the action of an electric field F acting along the z-direction. Electrons were set to move in the energy sites centered around LUMO and holes around HOMO. Carriers are allowed to hop within a cube consisting of 7x7x7 sites. The conventional Miller-Abrahams expression[19] has been used for the rate of hoping of a carrier from a site with energy of $\varepsilon_i$ to a site with an energy $\varepsilon_j$ at a distance $R_{ij}$,

$$\nu_{ij} = \nu_0 \exp\left(-\frac{\gamma \Delta R_{ij}}{a}\right) \begin{cases} \exp(-\frac{\varepsilon_j - \varepsilon_i}{kT}); \varepsilon_j > \varepsilon_i \\ 1; \varepsilon_j < \varepsilon_i \end{cases} \qquad - (2)$$

Here $R_{ij}=|R_i-R_j|$ and $\gamma$ is the overlapping factor between the two lattice sites which has been taken to be equal to *5/a*. Electrostatic potential (*eFx*) generated by a field is incorporated in the site energies $\varepsilon_i$ and $\varepsilon_j$.

The probability ($P_{ij}$) that a carrier jumps from a site *i* to any site *j* within a cube consisting of 7x7x7 sites is

$$P_{ij} = \frac{v_{ij}}{\sum_{i \neq j} v_{ij}}$$

- (3)

A random number from a uniform distribution is chosen and this specifies to which site the particle jumps because each site is given a length in random number space according to $P_{ij}$. The time for the jump is determined from

$$t_{ij} = x_{ei} \left[ \sum_{i \neq l} v_{il} \right]^{-1}$$

- (4)

Where $x_{ei}$ is taken from an exponential distribution of random number.

To include the effect of difference of mobility of electron and holes in the emissive material, hops of carriers are made time dependent, i.e. if one carrier takes longer time for a jump than the other carrier, jump of that carrier has not been made(while the other carrier is moving) until the time becomes the same for both the carriers. Time dependent hops can be explained by the following example:

Let electron and holes are at $z=Z_1 a$, $z=Z_2 a$ planes at a particular instant of time (assume t=0). Let us assume that for one jump hole takes a time $t_0$ and electron $10t_0$, where $t_0$ is an unit time. Therefore after a time $10t_0$, hole will make ten jumps while electron will make only one jump. Therefore the average electron will be found at $z= (Z_1+1) a$ plane where as holes will be found at $z= (Z_2-10) a$ plane, here it was assumed that electron is moving in +z direction and hole in –z direction and if jumps to neighbouring sites are assumed. To account this effect in the computation after every hop the time scale for electron and hole has been checked and slow carriers are stopped after a hop whereas faster carrier continues to make hops until the time scale for both the carriers become the same.

The computation was terminated when (i) A negative carrier reached at z=Za plane or positive carrier reached the z=1a plane (ii) a negative carrier and a positive carrier reach very close to each other (less than 2nm) where they recombine and a photon of energy equal to the difference of two energies is generated.

The later condition ignores dissociation of an exciton, equivalent to an infinite sink approximation. We have recorded the energy of emitted photons and the number of photons emitted at different energies. EL spectrum is generated for the device by plotting the number of photons (y-axis) against energy of photon (x-axis).

To set up the simulation, first we have taken the parameters (HOMO, LUMO) of a commonly used organic solid $Alq_3$. Curve a of Fig. 1(a) shows the experimental EL spectrum of an OLED with $Alq_3$ as an emissive layer having a device structure ITO (120 nm) / -NPD (300 nm) /$Alq_3$ (350 nm) /LiF (1 nm) /Al (150 nm). OLEDs were fabricated by the experimental procedure given in previous works[3]. It can be seen from the experimental curve that peak luminescence comes at a wavelength of 525 nm with a broad spectrum having Full width at half maximum (FWHM) of 90 nm. Now we have simulated the EL spectrum for $Alq_3$ using Monte-Carlo technique described above. The used values of HOMO-LUMO parameters are $E_{HOMO}$=5.7 eV, $E_{LUMO}$=3.0 eV. To fit the experimental EL spectrum, variance of LUMO distribution has been varied keeping the variance of HOMO distribution fixed at 125 meV. The parameters, which were fitted, are the wavelength of peak luminescence and FWHM of EL spectrum. It was found that the wavelength of peak luminescence does not depend on the variance of LUMO distribution ($\sigma_{LUMO}$) while FWHM varies with the variation of variance. Hence variance was adjusted in order to match the FWHM of EL spectrum. The value of $\sigma_{LUMO}$ at which the FWHM of simulated spectrum fit with experimental spectrum was 100meV. Curve b of Fig.1(a) shows the simulated EL spectrum with $\sigma_{LUMO}$=100 meV. Peak luminescence comes at 460 nm (which is invariant with variance) with a broad spectrum having a FWHM of 90 nm. Hence by adjusting the value of the variance, we were able to fit the FWHM but not the wavelength of peak EL. The reason for this disagreement may be that holes and electrons relax to tail states while hoping around HOMO and LUMO energy states. Due to this relaxation of carriers, peak of carrier distribution shifts to lower energies and hence the energy corresponding to peak emission is less than the HOMO-LUMO gap. Hence this point must be included into the technique. Previous studies on dielectric diffusion and relaxation of energy in disordered organic solids point out that a particle started at arbitrary energy within a Gaussian DOS of an undiluted system of hopping sites is likely to execute a random walk and relax into tail states[20]. It is a feature of a Gaussian DOS that the mean energy, saturates at long times indicating attainment of dynamic equilibrium. Typically, for the case of diffusion, relaxation of energy will take place in infinite time[12]. But when an external field is applied it creates a gradient in the relative position of the energy states thus creating a preferred direction and breaking the symmetry[13]. As typically a jump downwards in energy is favorable to a jump upwards, most of the carriers are relaxed in the tail states of Gaussian energy DOS while moving. Concept of transport energy ($\varepsilon_t$) has been used to explain the energy relaxation of a carrier in Gaussian DOS[20, 21] previously. Carriers in energy levels above $\varepsilon_t$ fall down in energy via hops to spatially neighboring states with lower energies. At $\varepsilon_t$ the character of relaxation changes. After a hop to a state below $\varepsilon_t$, the carrier prefers to hop upward towards $\varepsilon_t$. These hopping process near an below $\varepsilon_t$ resemble a multiple-trapping-like process, where $\varepsilon_t$ plays the role of mobility edge. The relaxation time is very low as compared to the movement of charge carriers in a Gaussian DOS. Hence the charge carriers can be assumed to be moving in a Gaussian profile centered at $\varepsilon_{tLUMO}$ below to $E_{LUMO}$ for electron and $\varepsilon_{tHOMO}$ above to $E_{HOMO}$ for holes due to relaxation of carriers. Now, this relaxation of charge carriers has been included in the simulation technique. The values of transport energy have been calculated by using the formalism given by Arkhipov et. al[22] for Gaussian DOS using the expression

$$\int_{Es}^{\varepsilon_t} dE_t g(E_t)(\varepsilon_t - E_t)^3 = \frac{6}{\pi}(\gamma kT)^3 \qquad - (5)$$

where $E_s$ is the energy of starting hopping site, $E_t$ is the energy of target site and $\varepsilon_t$ is the effective transport energy.

Movement of a charge carrier has been assumed to be in a Gaussian profile centered at energy $E_{LUMO} - \varepsilon_{tLUMO}$ (for negative carriers) and $E_{HOMO} + \varepsilon_{tHOMO}$ (for positive carriers). Wavelength corresponding to the peak luminescence becomes sensitive to the variation of variance of energy distribution. And after such a definition, electrical bandgap is less than the HOMO-LUMO gap by an amount $\varepsilon_{tLUMO} + \varepsilon_{tHOMO}$. Fig. 1(curve c) shows the simulated EL spectrum of Alq$_3$ after including the relaxation of carriers into the simulation. Fitting is excellent for the same parameters as used for fitting of FWHM without including the relaxation $\sigma_{HOMO}$=125 meV and $\sigma_{LUMO}$=100 meV. For clearity, the intensity values of curve c have been offseted by 0.3. One of the advantages of establishing simulation technique for EL spectrum is that one can change the system parameters separately at will in order to assess their influence on the system irrespective of experimental constraints. One of the parameter of interest is energetic disorder caused by the Gaussian energy distribution. Further, transport energy has been simulated for different values of variance using Eq. (5). Figure 1(b) shows the variation of transport energy in units of as a function of temperature for different values of variances 25, 50, 75 and 100 meV. It can be seen from the figure that transport energy is a function of temperature and variance.

After including the relaxation time of carriers, the peak of EL spectrum becomes dependent on variance of energy distribution. In an OLED, the disorder can be controlled by substrate temperature during the vacuum deposition. The higher the substrate temperature, the lower will be the disorder. Therefore, we have fabricated the OLED at different substrate temperatures 297 K, 347 K and 397 K. The peak of EL spectrum shifts from 525 nm to 505 nm by increasing the substrate temperature from 297 K to 397 K. The peak of EL spectrum for the OLED fabricated at 347 K was found to be at 516 nm. This can be explained by using the results of simulation. As the substrate temperature increases, the variance decreases, thereby, decreasing the value of the transport energy and $\varepsilon_{tLUMO} + \varepsilon_{tHOMO}$. Due to this reason, the difference between band gap and the energy corresponding to EL peak decreases or the EL spectrum is blue shifted.

Transport energy calculated for disordered solids, is dependent on temperature, the dependency of transport energy on temperature gives rise to temperature dependent EL spectrum. As the temperature increases, transport energy decreases and hence $\varepsilon_{tLUMO} + \varepsilon_{tHOMO}$ decreases. Due to this the difference between the optical band gap and energy corresponding to peak EL intensity decreases. This gives rise to temperature dependent EL spectrum. To fit our simulated spectrum with experiment, we have measured the EL spectrum at 200, 225, 250, 275 and 300 K. The EL spectrum was found to be red shifted with the decrease in temperature. This can be ascribed as due to the fact that the transport energy increases with the decrease in temperature as can be seen in Fig. 1(b) and therefore $\varepsilon_{tLUMO} + \varepsilon_{tHOMO}$ increases and the EL spectrum shifts towards the lower energy.

Further, a set of electroluminescent devices have been fabricated having a device structure ITO/ -NPD/EML/LiF/Al using Liq, Alq$_3$, Zn(hpb)mq as emissive layers (EML) to study the effect of energetic disorder. These molecules were selected due to their nearly matching values of HOMO-LUMO levels. For Liq LUMO and HOMO are centered at energies 3.1, 5.7 eV; 3.0, 5.7 eV for Alq$_3$ and 3.0, 5.8 eV for Zn(hpb)mq. Hence the HOMO-LUMO gaps are 2.6, 2.7, 2.8 eV for Liq, Alq$_3$ and Zn(hpb)mq respectively. In accordance to the HOMO-

LUMO gap, the wavelength corresponding to peak emission is expected to be the lowest for Zn(hpb)mq and the highest for Liq. Thus the red shift of wavelength is expected in the EL spectrum of Liq in comparison to the EL spectrum of Zn(hpb)mq. Figure 2 shows the recorded EL spectrum for the OLEDs with Liq, Alq$_3$, Zn(hpb)mq as emissive layers. Peak EL emission comes at 495, 525 and 550 nm for OLEDs with Liq, Alq$_3$ and Zn(hpb)mq respectively. The results of experimental EL observations were found to be just opposite to the expected results. Now simulation has been used to fit the experimental EL spectrum for these three devices. Reported HOMO-LUMO values of these materials have been used for simulation. For comparison, the variance of HOMO energy distribution was kept fixed at 125 meV except for Liq because for Liq the difference between the HOMO-LUMO gap and the energy corresponding to peak emission is 0.1 eV while the transport energy for 125 meV variance is 0.2 eV. The variance of LUMO distribution has been varied in order to fit the experimental EL spectrum. Excellent fits between the experimental and simulated EL spectrum were achieved for the value of variance of LUMO distribution 50, 100, 150 meV for Liq, Alq$_3$ and Zn(hpb)mq respectively, for Liq the fitted value of HOMO variance was 80 meV. According to these values the peak of negative carrier distribution profile is shifted by 0.036, 0.12, 0.35 eV below $E_{LUMO}$ for Liq, Alq$_3$, Zn(hpb)mq respectively, while the peak of positive carrier distribution is shifted by same amount for Alq$_3$ and Zn(hpb)mq (0.2 eV) and 0.06 eV for Liq due to the change of transport energy with variance. Variance causes different shift in the peak carrier density distribution in different materials. Due to this the actual difference between peak negative and positive carrier distribution is 2.5, 2.38, 2.25 eV for Liq, Alq$_3$ and Zn(hpb)mq respectively. This causes the red shift of EL spectrum of Zn(hpb)mq in comparison to the EL spectrum of Liq. Also from the experimental EL spectrum it can be seen that FWHM of EL spectrum is different for the three devices. FWHM value is highest for the EL spectrum of Zn(hpb)mq and lowest for Liq. This can be explained on the basis of the increasing value of variance from Liq to Zn(hpb)mq. The value of variance of LUMO distribution is highest for Zn(hpb)mq and lowest for Liq. Due to this the LUMO energy states of Zn(hpb)mq are more disordered in comparison to the LUMO energy states of Liq. This causes the broadening of the EL spectrum of Zn(hpb)mq in comparison to the EL spectrum of Liq. Hence from the results of simulation applied on the experiments, we can say that the EL spectrum observed from the disordered organic solids is highly dependent on variance of energy distribution.

The temperature dependent EL spectrum measurement has also been performed for the devices with Liq and Zn(hpb)mq. The shift in the EL spectrum for the Liq device was found lower in comparison to the shift in EL spectrum for the Zn(hpb)mq device with temperature. This can be ascribed as due to the lower value of variance in case of Liq in comparison to the Zn(hpb)mq and it can be seen from the Fig. 1(b) that the shift in transport energy with temperature is less in case of lower variance in comparison to the higher variance.

Further, the analytical treatment was carried by considering the charge carrier recombination in organic materials to be a bimolecular reaction, in which, the generation rate is given by

$$G = \frac{1}{4}\gamma n p \qquad (6)$$

where $\gamma$ is recombination constant and n, p are the negative and positive charge carrier concentrations. By assuming the charge carrier recombination to be a Langevin – type recombination, $\gamma$ can be related to charge carrier mobility $\mu$

$$\frac{\gamma}{\mu} = \frac{e}{\epsilon \epsilon_0} \qquad (7)$$

Therefore, $\gamma$ is dependent on mobility and mobility is dependent of charge carrier relaxation in disordered system, thereby making the generation rate to be dependent on charge carrier relaxation. The effect of charge carrier relaxation can be included in charge carrier mobility by using variable – range hopping model [20-22] and considering the carrier relaxation around the effective transport energy level. Further, using the Einstein relation yields following expression of equilibrium mobility [22]

$$\mu = \frac{ev_0}{kT}\left[\int_{-\infty}^{\infty} dE g(E) \exp\left(\frac{-E}{kT}\right)\right]^{-1} \left[\int_{-\infty}^{E_{tr}} dE g(E)\right]^{1/3} \exp\left(\frac{-E_{tr}}{kT}\right) \qquad (8)$$

Equation (6), (7) and (8) makes the photon generation rate to be dependent on transport energy. Now, the energy corresponding to peak generation rate can be evaluated simply equating

$$\frac{dG}{dE} = 0 \qquad (9)$$

Equation (6-9) yields the following expression for the energy of peak emission

$$\frac{1}{3}g(E_{tr}) \int_{-\infty}^{\infty} dE_m g(E_m) exp\left(\frac{-E_m}{kT}\right) - g(E_m) exp\left(\frac{-E_m}{kT}\right) \int_{-\infty}^{E_{tr}} g(E_m) dE_m = 0 \qquad (10)$$

where $E_m$ is the energy of peak emission and g(E) is DOS given by eq. (1). The spectral shift can then be simply evaluated by

$$E = E_g - E_m \qquad (11)$$

where $E_g$ is the HOMO-LUMO gap.

Transport energy varies with temperature. Therefore, we have evaluated the transport energy for Alq$_3$, Liq and Zn(hpb)mq at different temperatures using eq. 5 and plotted the experimental spectral shift obtained at different temperatures as a function of transport energy in Fig. 3. Analytical plot of spectral shift as a function of transport energy was also obtained using eq. 9 and 10 and shown in Fig. 3. Three different plots correspond to values used earlier for the simulation. It can be seen from the figure that the analytical results matches vary closely to experimental results which supports the applicability of our analytical treatment of spectral shift.

We conclude that the energetic disorder in Gaussian type density of state distribution affects the peak and FWHM of EL spectrum. Many puzzling experimental disagreement of EL such as the peak of EL occurring at lower energies than HOMO-LUMO gap, shifting of EL peak for materials with almost same HOMO-LUMO gaps by a large amount has been answered by setting up simulation for EL spectrum. An analytical treatment has been performed to support the results.

**Figure Captions:**

Fig. 1: (a) curve a shows the experimental EL spectrum for the OLED with $Alq_3$ as emissive layer, curve b shows the simulated EL spectrum without relaxation and curve c shows the simulated EL spectrum with relaxation, the intensity values of curve c has been offseted by 0.3 for clarity. (b) Simulated transport energy as a function of temperature for different variances.

Fig. 2: Experimental and simulated EL spectrum for Liq, $Alq_3$ and Zn(hpb)mq. Filled symbols are experimental curves while the open symbols are simulated curves. Intensity values of simulated curves are off sated by 0.3 for in comparison to their experimental curve for clarity.

Fig. 3: Experimental and analytical spectral shift as a function of transport energy, lines are the analytical results.

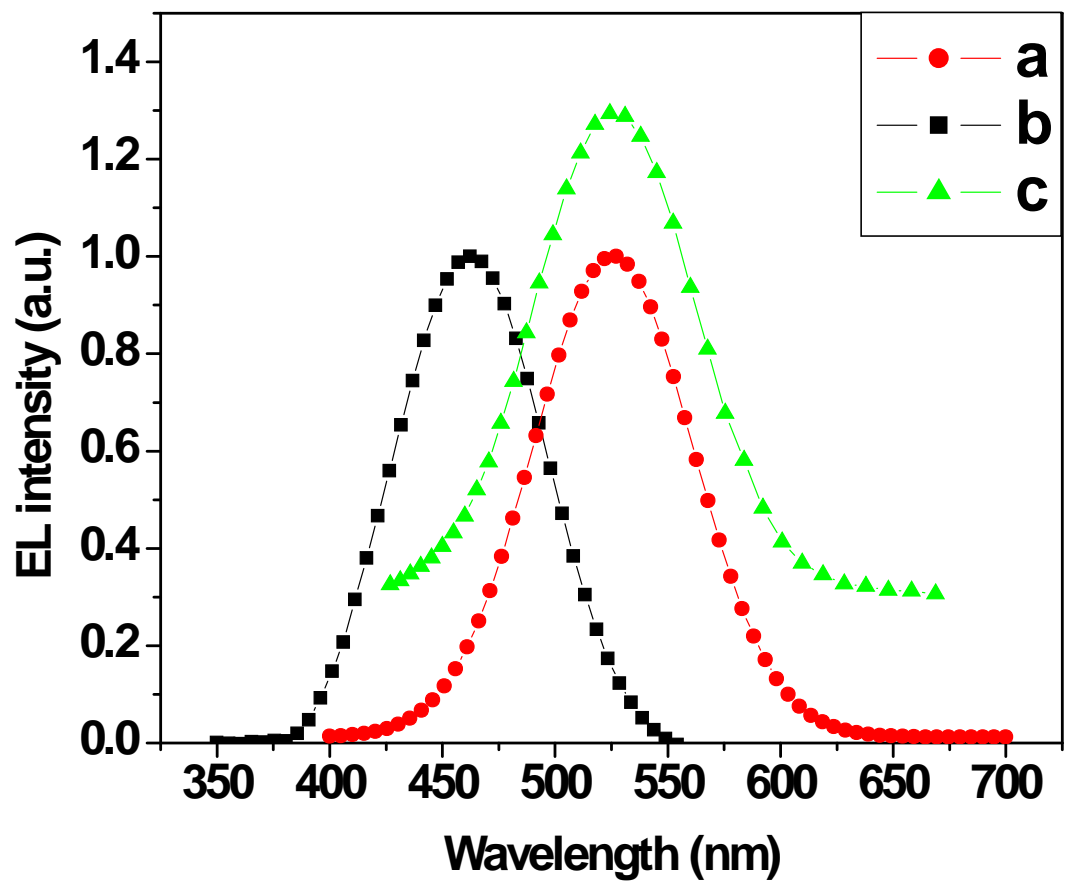

(a)

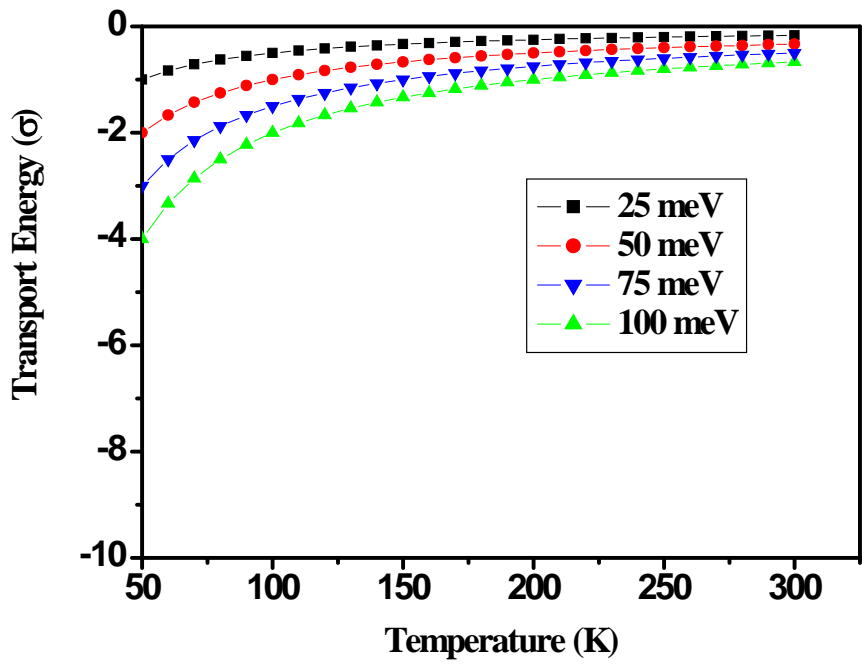

(b)

Fig. 1

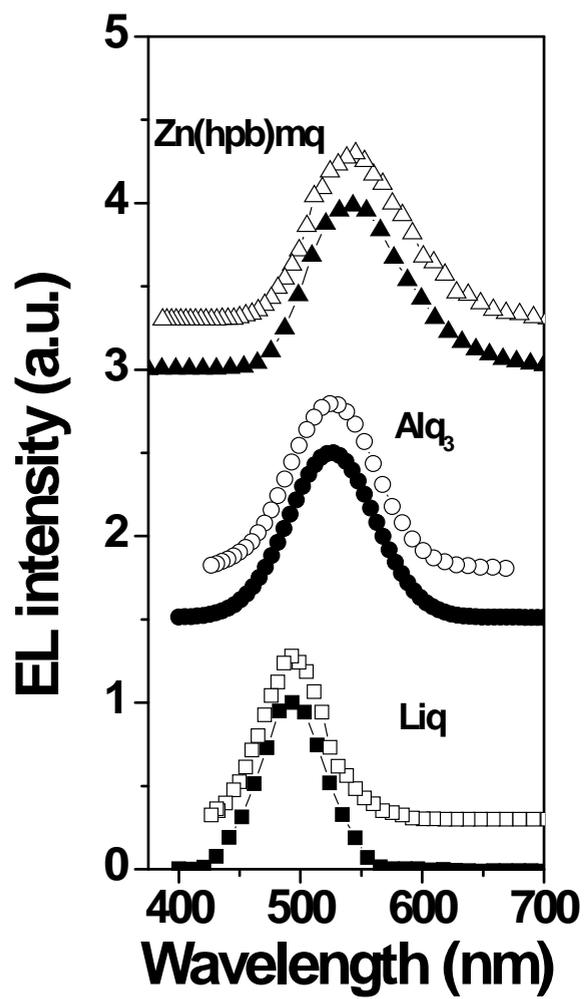

Fig. 2

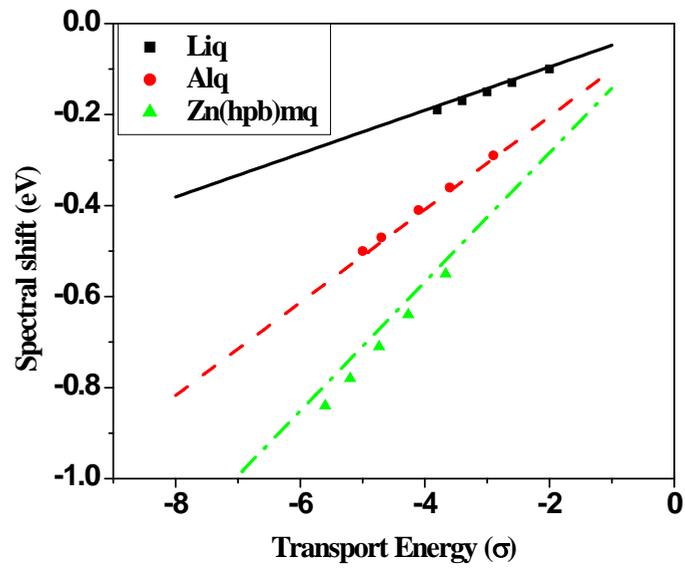

**Fig. 3**